\documentclass[aps,prb,twocolumn,showpacs]{revtex4}
\usepackage{graphicx}
\usepackage{epsfig}
\usepackage{amsfonts}
\usepackage{amscd}
\usepackage{bm}
\usepackage{amsmath}
\usepackage{amssymb}
\def\nicefrac#1#2{\frac{#1}{#2}}
\usepackage{mathptmx}
\usepackage{color}
\usepackage[dvips,breaklinks=false,colorlinks=true,urlcolor=slateblue,frenchlinks=false,bookmarks=false,pdfpagemode=None]{hyperref}
\DeclareGraphicsExtensions{.eps, .pdf, .jpg, .tif}
\usepackage{bookmath}
\usepackage{mathfrak}
\usepackage{mymatrices}
\usepackage{mycolors}
\begin{document}
\title{
\vspace*{-20mm}
\hspace*{-10cm}{\small Published in Phys. Rev. B. 84, 214509 (2011) 
[\href{http://dx.doi.org/10.1103/PhysRevB.00.004500}{doi: 10.1103/PhysRevB.00.004500}].
}\\
\vspace*{5mm}
Surface States, Edge Currents and the Angular Momentum of Chiral p-wave Superfluids}
\author{J.A. Sauls} 
\affiliation{Department of Physics and Astronomy, 
             Northwestern University, Evanston, Illinois 60208}
\date{November 11, 2011}
\begin{abstract}
The spectra of fermionic excitations, pairing correlations and edge currents confined near the
boundary of a chiral p-wave superfluid are calculated to leading order in $\hbar/p_f\xi$.
Results for the energy- and momentum-resolved spectral functions, including the spectral current
density, of a chiral p-wave superfluid near a confining boundary are reported. The spectral
functions reveal the subtle role of the chiral edge states in relation to the edge current and the
angular momentum of a chiral p-wave superfluid, including the rapid suppression of $L_z(T)$ for $0
\lesssim T\ll T_c$ in the fully gapped 2D chiral superfluid.
The edge current and ground-state angular momentum are shown to be sensitive to boundary
conditions, and as a consequence the topology and geometry of the confining boundaries.
For perfect specular boundaries the edge current accounts for the ground-state angular momentum,
$L_z=(N/2)\hbar$, of a cylindrical disk of chiral superfluid with $N/2$ fermion pairs.
Non-specular scattering can dramatically suppress the edge current.
In the limit of perfect retro-reflection
the edge states form a flat band of zero modes that are non-chiral and generate no edge current.
For a chiral superfluid film confined in a cylindrical toroidal geometry the ground-state angular
momentum is, in general, non-extensive, and can have a value ranging from $L_z > (N/2)\hbar$
to $L_z < -(N/2)\hbar$ depending on the ratio of the inner and outer radii and the degree of
back scattering on the inner and outer surfaces.
Non-extensive scaling of $L_z$, and the reversal of the ground-state angular momentum for a
toroidal geometry, would provide a signature of broken time-reversal symmetry of the ground
state of superfluid \Hea, as well as direct observation of chiral edge currents.
\end{abstract}
\maketitle
\vspace*{-7mm}
\subsection{Introduction}
\vspace*{-5mm}

Among the remarkable phases of liquid \He\ is the A-phase. In addition to being a superfluid which
supports persistent currents, this fluid is believed to possess a spontaneous mass current in its
ground state.
Ground state currents are associated with the chirality of Cooper pairs that
condense to form the A-phase and conspire to produce a macroscopic angular momentum.
Chirality is encoded in the p-wave orbital order parameter, $\Delta(\vp)
=\Delta\,\vp\cdot\left(\hvm+i\hvn\right)/p_f =\Delta\sin\theta_{\vp}\,e^{i\phi_{\vp}} $, where $\vp$
is the relative momentum of a Cooper pair, $\{\hvm,\hvn,\hvl\}$ is an orthonormal triad of unit
vectors that define the orbital coordinates of the Cooper pair wave function, and
$\Delta\sim k_B T_c$ is the pairing energy.\cite{vollhardt90}
This order parameter is an eigenfunction of the orbital angular momentum along $\hvl=\hvm\times\hvn$
with eigenvalue $+\hbar$.
Such broken symmetries in bulk condensed matter systems have implications for the spectrum of
excitations bound to surfaces and topological defects.\cite{volovik92}
This phase breaks time-reversal symmetry as well as parity, and is realized at all pressures below
melting in thin superfluid $^3$He-A films. In the 2D limit the Fermi surface is fully gapped, and
belongs to the topological class of integer quantum Hall systems.\cite{rea00,vol88,vol92}
The 2D A-phase is also representative of layered p-wave superconductors with broken time-reversal 
symmetry, e.g. the proposed order parameter for superconducting \sro.\cite{mac03}

The macroscopic manifestation of chiral order in \Hea\ is the ground-state angular momentum,
$\vL = \int_{V}dV\vr\times\vg(\vr)$, where $\vg$ is the mass current density. For 2D chiral
p-wave superfluids in the BCS limit, where the size of Cooper pairs is large compared to Fermi
wavelength, $\xi\gg\hbar/p_f$, the ground state currents are predominantly confined to boundaries.
I discuss effects of surface scattering on the pairing correlations, the fermionic spectrum and
ground state currents in the vicinity of boundaries confining a chiral p-wave superfluid.
Results for the spectral current density highlight the fermionic spectrum that is
responsible for the edge current and the ground state angular momentum.
The theory is extended to finite temperatures, non-specular boundaries and multiply connected 
geometries.
The results reported here are discussed in context with the results of Kita,\cite{kit98} and 
Stone and Roy.\cite{sto04}

Starting from Bogoliubov's equations in Sec. \ref{Bogoliubov-Andreev-Eilenberger}, I introduce
Eileberger's quasiclassical equation for the Nambu propagator that is the basis for investigating
the pairing correlations, spectrum of surface states and edge currents for chiral p-wave
superfluids. 
The bound-state spectrum and results for the spectral current density are discussed in
Secs. \ref{sec:chiral_edge_state} and \ref{sec:spectral_current_density}. 
Analysis of the continuum spectrum, edge current and the spectral analysis of the ground state
angular momentum are reported in Sec. \ref{sec:edge_current},
which is followed by results and a discussion of the temperature
dependence of the edge current and angular momentum in Sec. \ref{sec:Lz_vs_T}. 
In Sec. \ref{sec:boundary_conditions} I discuss the sensitivity of the edge current and ground-state
angular momentum to boundary scattering and geometry, and in Sec. \ref{sec:toroidal_geometry}
discuss the non-extensive behavior of the ground-state angular momentum that develops for multiply
connected geometries in which there is an asymmetry in the specularity on different surfaces.
I start with some background on the ground state current and angular momentum of superfluid \Hea.

The magnitude of the ground-state angular momentum, $L_z$, has been the subject of considerable
theoretical investigation.
Predictions for $L_z$ of \Hea\ in a cylindrically symmetric vessel vary over many orders of
magnitude,\cite{and61,vol75,cro77,ish77,leg78,mcc79} from $L_z\simeq(N/2)\hbar\,(\Delta/E_f)^2$ to
$L_z=(N/2)\hbar$, where $N/2$ is the total number of fermion pairs in the volume $V$.
This latter result is what one intuitively expects for Bose-Einstein condensate (BEC) of
tightly bound molecules, each carrying one unit of angular momentum, and whose molecular
size $\xi$, is small compared to the mean distance between molecules,
$a\equiv\sqrt[3]{\mbox{\small $2V/N$}}\gg\xi$.
However, this result is also obtained in the opposite limit, $\xi\gg a$, appropriate to
BCS condensation of Cooper pairs, each with angular momentum $\hbar\,\hvl$ and radial size
$\xi=\hbar v_f/\pi\Delta$, where one expects almost exact cancellation of the internal
currents from overlapping Cooper pairs.\cite{ish77}
In particular, McClure and Takagi \cite{mcc79} showed that an N-particle ground state of
the form,
\ber\label{BCS_wave-function}
\ket{N}=\left[\iint\,d\vr d\vr'
              \varphi_{\alpha\beta}(\vr,\vr')\,
              \psi^{\dagger}_{\alpha}(\vr)\psi^{\dagger}_{\beta}(\vr')
        \right]^{N/2}
\hspace*{-5mm}\ket{\mbox{vac}}
\,,
\eer
with an equal-spin, odd-parity chiral pairing amplitude,\footnote{I will refer to this order
parameter as the `Anderson-Morel (AM) state', the `chiral p-wave state' or the `A-phase order parameter'.}
\ber
\cF_{\alpha\beta}
&\equiv&
\bra{N-2}\psi_{\alpha}(\vr+\vx/2)\psi_{\beta}(\vr-\vx/2)\ket{N}
\nonumber\\
&=&\cF(|\vr|)\,
\vec{\vd}\cdot(i\vec\vsigma\sigma_y)_{\alpha\beta}\,
\left(\hvm(\vr) + i \hvn(\vr)\right)\cdot\vx
\,,
\eer
of the AM form that preserves cylindrical symmetry is an eigenstate of the
total angular momentum with $L_z = (N/2)\hbar$.\footnote{This result is also applicable
to spatially varying ``textures'' of the AM state in which $\hvl(\vR)$ varies slowly on the scale
of $\xi$, such as the Anderson-Brinkman \cite{and75} and Mermin-Ho \cite{mer76} textures for \Hea\
in a long cylinder.}
Thus, the ground-state angular momentum of a chiral condensate is the same for $N/2$ Bose molecules
or $N/2$ Cooper pairs. However, the magnitude and spatial distribution of the mass currents that
give rise to the total angular momentum differ in the BEC and BCS limits.
This somewhat non-intuitive result is intimately connected
with the symmetry of the ground state and its implications for the surface fermionic spectrum and
associated currents.\cite{vol97,sto04}
Numerous authors have addressed the question of the current distribution responsible for the ground
state angular momentum.\cite{ish77,cro77,mer80,bal85}$^{,}$\footnote{See Volovik and Mineev's review
of this subject in Ref. \cite{vol81b}, and also for a clear discussion of differences between chiral
Bose molecules and Cooper pairs in \Hea.}
Starting from the N-particle BCS wavefunction in Eq. \ref{BCS_wave-function},
Ishikawa\cite{ish77} and Mermin and Muzikar\cite{mer80} calculated the current density in
the long wavelength limit, $\cL \gg \xi$, for the AM state at $T=0$. For spatially
uniform $\hvl$ and no center of mass supercurrent,
\be\label{hydro_bound_current}
\vg = \curl{(\nicefrac{1}{4}\,n\,\hbar\,\hvl)}
\,.
\ee
In the BCS limit the density, $n(\vr)$, is spatially uniform \emph{except} near the boundary,
$\vr=\vR$, where $n(\vR)=0$. The current is then confined at the boundary, $\vg =
\nicefrac{1}{4}\hbar\,(-\partial n/\partial r)\,\hat\phi$, from which one recovers the result
for the ground-state angular momentum
$\vL=\int_{\text{V}}dV\,\vr\times\vg=(N/2)\hbar\,\hvl$.
%
This highlights a limitation of the gradient expansion and hydrodynamic limit. The order parameter
is assumed to be the local equilibrium the AM state, and spatial variations are assumed to be long
wavelength on the scale of $\xi\gg a$. However, the density varies on atomic length scales near the
surface, whereas the order parameter is, in general, strongly deformed on on length scales of order
$\xi$ near a boundary. Thus, Eq. \ref{hydro_bound_current}, and the gradient expansion in
particular, do not accurately describe the current density near the boundary, nor do they account
for the source of the surface current.
This requires a theory valid for spatial variations of the condensate on length scales
comparable to or smaller than the correlation length $\xi$.

\begin{figure}[t]
\vspace*{-5mm}
\includegraphics[width=0.9\linewidth,keepaspectratio]{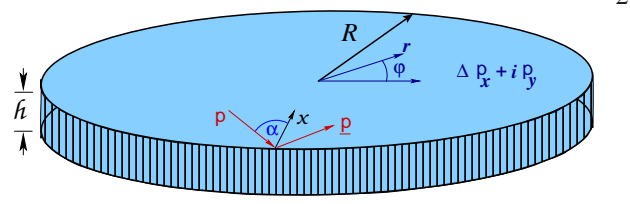}
\caption{\small
A thin film of $p_x+ip_y$ superfluid (``2D \Hea'') confined in a cylindrical geometry with
thickness $\mathit{h}\ll\xi$, radius $R \gg \xi$ bounded by specular surfaces which reflect
excitations, $\vp\rightarrow\underline{\vp}$.
\label{fig-cylindrical_film_geometry}
}
\end{figure}

\vspace*{-7mm}
\subsection{Bogoliubov-Andreev-Eilenberger}\label{Bogoliubov-Andreev-Eilenberger}
\vspace*{-5mm}

For a thin film of \Hea, as shown in Fig. \ref{fig-cylindrical_film_geometry}, the orbital
quantization axis is locked normal to the surface of the film, $\hvl || \hvz $.\cite{amb75}
The A-phase also belongs to the class of equal spin pairing (ESP) states with spin structure of the
order parameter given by a linear combination of the symmetric Pauli matrices,
\be\label{ESP_A-phase}
\Delta_{\alpha\beta}(\vp)=\vd\cdot(i\vec\sigma\sigma_y)_{\alpha\beta}\,\Delta(\vp)
\,,
\ee
where $\alpha,\beta$ label the projections of fermion spins of the Cooper pair
and $\vd$ is the direction in spin space along which Cooper pairs have zero spin projection.
Thus, for $\vd=\hvz$ the spin state of the Cooper pairs is given by
$i\vec\sigma\sigma_y\cdot\vd=\sigma_x$, which is the triplet state with equal amplitude
for the Cooper pairs to be spin polarized along $+\hvx$ or $-\hvx$:
$\ket{\vd}=\nicefrac{1}{\sqrt{2}}\left(\ket{\rightrightarrows} +
\ket{\leftleftarrows}\right)$.
Spin textures described by spatial variations of the $\vd$ vector are possible;
however, in what follows I assume the spin state is fixed by 
the nuclear dipolar energy which locks the $\vd\,||\, \hvl$.\cite{leg75}
The bulk A-phase of \He\ in 3D has gapless excitations for momenta along the nodal directions,
$\vp\, || \pm\,\hvl$. Here I consider 2D \Hea\ with a cylindrical Fermi surface, (or a set of
cylindrical Fermi surfaces generated by dimensional quantization), and an orbital order parameter
given by
\be\label{AM_order_parameter}
\Delta(\vp) = \Delta\left(p_x + i\,p_y\right)/p_f
\,,
\ee
which generates a bulk excitation spectrum that is fully gapped on the Fermi surface.

Near a boundary, or domain wall, the orbital order parameter can deviate from the pure A-phase form.
Thus, a more general form of the orbital p-wave order parameter is parametrized by two
real amplitudes,
\be
\Delta(\vr,\vp) = \Big(\Delta_{1}(\vr)\,p_x + i\,\Delta_{2}(\vr)\,p_y\Big)/p_f
\,,
\ee
with $\Delta_{1,2}(\vr)\rightarrow\Delta$ far from a boundary. 
Inhomogeneous states are described by the
Bogoliubov's equations\cite{vol88,kur90,vol97}
\ber
\,\left(-\frac{\hbar^2}{2m^*}\nabla^2 - \mu\right) u_{\alpha}(\vr)
 +\Delta_{\alpha\beta}(\vr,\vp)\,v_{\beta}(\vr) = \varepsilon\,u_{\alpha}(\vr)
\,,
\\
-\left(-\frac{\hbar^2}{2m^*}\nabla^2 - \mu\right) v_{\alpha}(\vr)
 +\Delta^{\dag}_{\beta\alpha}(\vr,\vp)\,u_{\beta}(\vr) = \varepsilon\,v_{\alpha}(\vr)
\,.
\eer
for the particle ($u_{\alpha}(\vr)$) and hole ($v_{\alpha}(\vr)$)
wavefunctions. For $\vd=\vz$ the Bogoliubov equations reduce to $2\times 2$
equations for Bogoliubov spinors, $\ket{\varphi}=\left(u,v\right)^{\text{T}}$, in Nambu (particle-hole) space,
\be
\widehat{\cH}_{\text{B}}\ket{\varphi} = \varepsilon\ket{\varphi}
\,,
\ee
where $\widehat{\cH}_{\text{B}}$ is the Bogoliubov Hamiltonian expressed in terms of Nambu
matrices, $\whtaux,\whtauy,\whtauz$,
\be
\widehat{\cH}_{\text{B}} 
= \xi(\vp)\whtauz 
+ \Delta_{1}(\vr,\vp)\whtaux
+ \Delta_{2}(\vr,\vp)\whtauy
\,,
\ee
with $\vp=\hbar/i\grad$, and the off-diagonal pair potentials interpreted as symmetrized operators,
\be
\Delta_{1,2}(\vr,\vp) 
= 
\frac{\hbar}{2i}
\left(
\Delta_{1,2}(\vr)\,\partial_{x,y} + \partial_{x,y}\, \Delta_{1,2}(\vr)
\right)
\,.
\ee
The large difference between the Fermi wavelength, $\hbar/p_f$, and the size of Cooper pairs, $\xi$,
is the basis for Andreev's quasiclassical approximation to the Bogoliubov equations.\cite{and64} The
expansion is achieved by factoring the fast- and slow spatial variations of the Bogoliubov spinor,
\be
\ket{\varphi} = e^{i\vp_f\cdot\vr/\hbar}\,\ket{\psi_{\vp_f}}
\,,
\ee
and retaining leading order terms in $\hbar/p_f\xi\ll 1$, which yields Andreev's equation,
\be\label{Andreev_Equation}
\widehat{\cH}_{\text{A}}\,\ket{\psi_{\vp}}
+
i\hbar\,\vv_{\vp}\cdot\grad\,\ket{\psi_{\vp}}
= 0
\,.
\ee
with operator $\cH_{\text{A}}$ defined by
\be
\widehat{\cH}_{\text{A}}=\varepsilon\whtauz - \whDelta(\vr,\vp)
\,,
\ee
and the Nambu matrix order parameter given is by
\be
\whDelta(\vr,\vp) = i\sigma_x\left(\whtaux\Delta_{2}(\vr,\vp) + \whtauy\Delta_{1}(\vr,\vp)\right)
\,,
\ee
where $\vp=p_f\hat{p}$ is the Fermi momentum, $\vv_{\vp}=v_f\hat{p}$ is the Fermi velocity. The
latter defines classical straight-line trajectories for the propagation of wavepackets of Bogoliubov
excitations, which are coherent superpositions of particles and holes with amplitudes given by the
Andreev-Nambu spinor,
\be\label{Andreev-Nambu_Spinor}
\ket{\psi_{\vp}}=\begin{pmatrix} {u}_{\vp} \cr {v}_{\vp} \end{pmatrix}
\,.
\ee
Andreev's equation expressed in terms of a row spinor is\footnote{The row spinors are not
simply the adjoint spinor of Eq. \ref{Andreev-Nambu_Spinor} since $\widehat{\cH}_{\text{A}}$ is not
Hermitian.}
\be
\bra{\tilde{\psi}_{\vp}}\widehat{\cH}_{\text{A}} -
i\hbar\vv_{\vp}\cdot\grad\,\bra{\tilde{\psi}_{\vp}} = 0
\,,
\ee
with the normalization of the Andreev-Nambu spinor given by
$\braket{\tilde{\psi}_{\vp}}{\psi_{\vp}}=1$.
There are two solutions (branches) to Andreev's equation for a single trajectory defined by $\vp$.
For $|\varepsilon|>\Delta$, the two branches are propagating solutions; a particle-like
solution, $\ket{\psi_{\vp\,+}}$, with group velocity $\vv(\varepsilon) || \vp$ and hole-like
solution, $\ket{\psi_{\vp\,-}}$, with reversed group velocity, $\vv(\varepsilon) || -\vp$.
For energies within the bulk gap the solutions are exploding and decaying amplitudes along the
trajectory, and thus relevant only in the vicinity of boundaries, domain walls, etc.

The product of the particle- and hole amplitudes in Eq. \ref{Andreev-Nambu_Spinor},
\be
\mff_{\alpha\beta}(\vr,\vp;\varepsilon) 
= {u}_{\alpha}(\vr,\vp;\varepsilon){v}_{\beta}(\vr,\vp;\varepsilon)
\,,
\ee
is the \emph{pair propagator}, which determines the spectral composition of the 
Cooper pair amplitude,
\be
\cF_{\alpha\beta}(\vr,\vp)
=
\int d\varepsilon\,f(\varepsilon)\,\,\mff_{\alpha\beta}(\vr,\vp;\varepsilon)
\,,
\ee
where $f(\varepsilon)=1/(e^{\varepsilon/T} +  1)$ is the Fermi distribution.
The pair propagator is one component of the Nambu matrix
\be
\whmfg(\vr,\vp;\varepsilon) = 
\sum_{\mu,\nu}\,g_{\mu\nu}\,\ket{\psi_{\vp\mu}}\bra{\tilde{\psi}_{\vp\nu}}
\,,
\ee
which satisfies Eilenberger's transport equation,\cite{eil68}
\be\label{Eilenberger_Equation}
\left[\widehat{\cH}_{\text{A}}\,,\,\whmfg(\vr,\vp;\varepsilon)\right] 
+ 
i\hbar\vv_{\vp}\cdot\grad\,\whmfg(\vr,\vp;\varepsilon) = 0
\,.
\ee
Physical solutions to Eq. \ref{Eilenberger_Equation} must also satisfy 
Eilenberger's normalization condition,\cite{eil68}
\be\label{Normalization_condition}
\left(\whmfg(\vr,\vp;\varepsilon)\right)^2 = -\pi^2\widehat{1}
\,.
\ee

An advantage of Eilenberger's formulation is that the spectral functions for both quasiparticle and
pair excitations are obtained as components of the quasiclassical propagator. For a fixed spin
quantization axis, $\vd=\hat\vz$, the off-diagonal components of the propagator describe pure
equal-spin pairing correlations. As a result the Nambu propagator can be expressed in the form,
\be\label{QCPropogator}
\whgR = \gR_{3}\whtauz + i\sigma_x\,\left(\fR_{2}\whtaux - \fR_{1}\whtauy\right)
\,.
\ee
The superscript refers to the causal (retarded in time) propagator,
obtained from Eq. \ref{Eilenberger_Equation} with the shift,
$\varepsilon\rightarrow\varepsilon+i0^{+}$.
%
The diagonal propagator in Nambu space, $\gR_{3}\whtauz$, determines the spectral function, or local
density of states, for the fermionic excitations with momentum $\vp=p_f\hat{p}$,
\be
\cN(\vr,\vp;\varepsilon) = -\frac{1}{\pi}\Im\,\gR_{3}(\vr,\vp;\varepsilon)
\,,
\ee
while the off-diagonal propagators, $\fR_{1}\whtauy$ and $\fR_{2}\whtaux$, determine the spectral
function for the correlated pairs,
\be
\cP_{1,2}(\vr,\vp;\varepsilon) = -\frac{1}{\pi}\Im\,\fR_{1,2}(\vr,\vp;\varepsilon)
\,.
\ee
These functions determine the mean pair potentials, $\Delta_{1}$
and $\Delta_{2}$, through the BCS self-consistency condition,
\ber
\Delta_{1,2}(\vr,\vp) 
=
\langle
v(\vp,\vp')
\int^{+\Omega_{\text{c}}}_{-\Omega_{\text{c}}}\hspace*{-5mm}d\varepsilon
\tanh\left(\frac{\varepsilon}{2T}\right)
\cP_{1,2}(\vr,\vp';\varepsilon)
\rangle_{\vp'}
\,,
\eer
where $\Omega_c\ll E_F$ is the bandwidth of attraction for the spin-triplet,
p-wave pairing interaction, $v(\vp,\vp')$, which is integrated over the occupied states defining the
pair spectrum and averaged over the Fermi surface,
$\langle\ldots\rangle_{\vp'}\equiv\int\,d\Omega_{\vp'}/4\pi(\ldots)$.

\vspace*{-7mm}
\subsection{Chiral Edge State}\label{sec:chiral_edge_state}
\vspace*{-5mm}

For a boundary far from other boundaries only single reflections, $\vp\rightarrow\ul{\vp}$, couple
the propagators for the incoming ($\vp$) and outgoing ($\ul{\vp}$) trajectories.
In particular, for the pair of specularly reflected trajectories on the boundary shown in Fig.
\ref{fig-cylindrical_film_geometry}, with radius of curvature large compared to the
correlation length, $R\gg\xi$, the solutions for the components of the propagator are (see Appendix
\ref{appendix-quasiclassical})
\ber
\fR_1(x,\vp;\varepsilon) 
  &=& \frac{\pi\Delta_{1}}{\lambda(\varepsilon)}\,
      \left(1 - e^{-2\lambda(\varepsilon)x/v_x}\right)
\,,
\label{fR1}
\\
\fR_2(x,\vp;\varepsilon) 
  &=& \frac{\pi\Delta_{2}}{\lambda(\varepsilon)}\,
   -  \frac{\pi\Delta_{1}}{\lambda(\varepsilon)}\,
\frac{\Delta_{1}^{2}-(\tepsR)^{2}}{\lambda\tepsR-\Delta_{1}\Delta_{2}}
\,e^{-2\lambda(\varepsilon)x/v_x}
\,,
\label{fR2}
\\
\gR_3(x,\vp;\varepsilon) 
  &=&
-\frac{\pi\tepsR}{\lambda(\varepsilon)}
+
\frac{\pi\Delta_{1}}{\lambda(\varepsilon)}\,
\frac{\tepsR\Delta_{1}+\lambda\Delta_{2}}
      {(\tepsR)^2-\Delta_{2}^{2}}
\,e^{-2\lambda(\varepsilon)x/v_x}
\,.
\qquad
\label{gR3}
\eer
where $v_x = v_f\cos(\alpha)$ for $-\pi/2<\alpha<\pi/2$ and $x\ge 0$ is the coordinate
normal to the boundary as shown in Fig. \ref{fig-cylindrical_film_geometry}.

Note that the propagator corresponding to Cooper pairs with relative momentum normal to the boundary
vanishes at the boundary, $\fR_{1}(x=0,\vp;\varepsilon)\equiv 0$. De-pairing of the normal amplitude
is partially compensated by an increase in the pairing correlations for pairs with relative momenta
parallel to the boundary. The origin of this enhancement is the fermionic state, bound to the
surface, which appears as a pole in the propagators of Eqs. \ref{fR2} and \ref{gR3} at the energies,
\be\label{chiral_bound-state}
\varepsilon_{\text{bs}}(\vp) = -\Delta_{2}(\vp) = - c\,p_{||}
\,.
\ee
The surface state disperses with momentum $p_{||}=p_f\sin\alpha$ parallel to the surface, 
$-p_f \le p_{||} \le +p_f$, and $c=\Delta/p_f\ll v_f$.

\begin{figure}[h]
\includegraphics[width=0.95\linewidth,keepaspectratio]{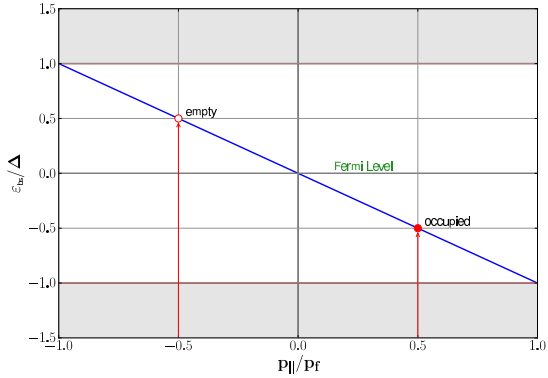}
\caption{\small
Chiral edge state dispersion, $\varepsilon_{\text{bs}}(\vp)=-c\,p_{||}$,
illustrating the asymmetry in the occupation of pairs of time-reversed states.
\label{fig-chiral_edge-state_dispersion}
}
\end{figure}

The important feature of the spectrum of surface fermions, shown in Fig.
\ref{fig-chiral_edge-state_dispersion}, is that there is \emph{no} branch with the opposite phase
velocity. The spectrum describes Weyl or chiral fermions.\cite{vol97}$^{,}$\footnote{There is a
two-fold spin degeneracy for each $\vp_{||}$.} For each pair of time-reversed fermions the state
with $+p_{||}$ is \emph{occupied} while its time-reversed partner with momentum $-p_{||}$ is
\emph{empty}. As a result the pairs of surface states generate a net mass or charge current. This
asymmetry in the occupation of the surface spectrum is a reflection of the chirality of the ground
state order parameter and specular reflection at the boundary which preserves translation symmetry
locally along the boundary.
The absence of a branch of fermions with energy $\varepsilon_{+}(\vp_{||}) = +\Delta_{2}(\vp)$ is
demonstrated by evaluating the residue of $\gR_3(x,\vp;\varepsilon)$ at the apparent pole,
$\varepsilon_{+}$: $\mbox{Res}\,\gR_3(x,\vp;\varepsilon)|_{\varepsilon_{+}}\equiv 0$. 
For energies in the vicinity of the bound-state pole,
$|\varepsilon-\varepsilon_{\text{bs}}| \ll |\Delta|$, the quasiparticle propagator reduces to
\be\label{gR_bound-state}
\gR_{\text{bs}}(x,\vp;\varepsilon) 
  =
\frac{\pi|\Delta_{1}(\vp)|}{\varepsilon +i\gamma - \varepsilon_{\text{bs}}(\vp)}
\,e^{-2\Delta\,x/v_f}
\,,
\ee
where I include the line-width, $\gamma\ll\Delta$, of the surface state due to weak disorder.
For $\gamma\rightarrow 0^{+}$ the states are sharp and the spectral function consists of
delta functions at $\varepsilon_{\text{bs}}(\vp)$,
\be\label{DOS_bound-state}
\cN_{\text{bs}}(x,\vp;\varepsilon) =\pi|\Delta_{1}(\vp)|\,e^{-2\Delta\,x/v_f}\,
                                    \delta(\varepsilon - \varepsilon_{\text{bs}}(\vp))
\,.
\ee
The spectral weight is maximum for trajectories at normal incidence and vanishes for grazing
incidence. Note that every edge state is confined to the surface on the length scale
\be
\xi_{\text{$\Delta$}} = \hbar v_f/2\Delta
\,,
\ee
independent of momentum $\vp_{||}$, and of order the Cooper pair size, $\xi_{\text{$\Delta$}} =
\hbar v_f/2\Delta \simeq 1.6\,\xi$.

\vspace*{-7mm}
\subsection{Spectral Current Density}\label{sec:spectral_current_density}
\vspace*{-5mm}

\begin{figure}[h]
\includegraphics[width=0.85\linewidth,keepaspectratio]{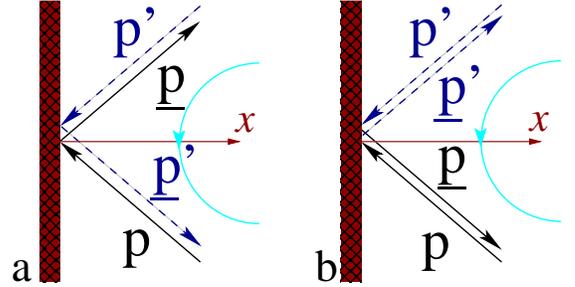}
\caption{\small
a) Time-reversed trajectory pairs that define the spectral current density,
$\cJ(x,\vp;\varepsilon)$, for specular reflections. b) Retro-reflections are time-reversed partners
for any incident angle. The chirality of the bulk order parameter is indicated by the direction of
the arc.
\label{fig-time-reversed_trajectories}
}
\end{figure}

The spectral current density is defined as the local density of current carrying states in the
energy interval $(\varepsilon,\varepsilon+d\varepsilon)$ for states with momentum $\vp$,
\be\label{JDOS}
\cJ(x,\vp;\varepsilon) = 2 N_f \vv_{\vp}\,
                         \left[
                         \cN_{\text{in}}(x,\vp;\varepsilon) -
                         \cN_{\text{in}}(x,\vp';\varepsilon)
                         \right]
\,,
\ee
where $\cN_{\text{in}}(x,\vp;\varepsilon)$ is the spectral function calculated for the incident
trajectory with momentum $\vp$, $N_f$ is the normal-state density of states at the Fermi level for
one spin, and $\vp$ and $\vp'$ define the pair of time-reversed incident trajectories shown in Fig.
\ref{fig-time-reversed_trajectories}a, for which $\vv_{\ul{\vp}'}=-\vv_{\vp}$.

The resulting local current density is obtained by thermally occupying the spectrum and 
integrating over all incoming trajectories,
\be\label{particle_current_density}
\vj(x) = 
\int_{\text{in}}\frac{d\Omega_{\vp}}{4\pi}\int_{-\infty}^{+\infty}\,d\varepsilon\,f(\varepsilon)\,
               \cJ(x,\vp;\varepsilon)
\,,
\ee
where $f(\varepsilon)=1/(e^{\varepsilon/T} + 1)$ is the Fermi distribution.

\begin{figure}[h]
\includegraphics[width=0.95\linewidth,keepaspectratio]{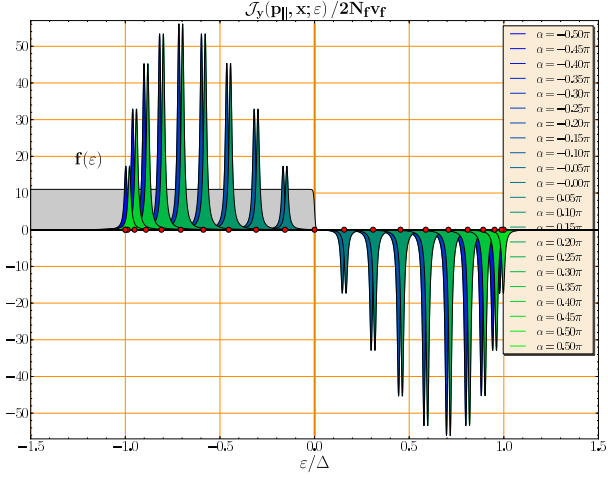}
\caption{\small
Spectral current density $\cJ_y$ for $x=0$ as a function of $p_{||}=p_f\sin\alpha$ for
linewidth $\gamma=0.025\Delta$.
States with $\pm p_{||}$ are slightly offset to show the contributions to the
current from time-reversed pairs.
\label{fig-Jydos_bound-state}
}
\end{figure}

The spectral current density for the bound-state spectrum obtained from Eqs. \ref{gR_bound-state}
and \ref{JDOS} is shown in Fig. \ref{fig-Jydos_bound-state} for the full range of incident
trajectories. Note that time-reversed states, incident angles $\alpha$ and $-\alpha$, add coherently
to the spectral current density. Thus, the net current density parallel to the boundary carried by
the surface bound states is given by\footnote{By contrast the component of the current \emph{into}
the boundary, $j_x^{\text{bs}}$, vanishes identically, as required by particle conservation.}
\ber
j_y^{\text{bs}}(x) 
&=&
I(\Delta/T) \times N_f\, v_f\, \Delta\, e^{-x/\xi_{\text{$\Delta$}}}
\,,
\label{bound-state_current-density}
\eer
where the integration over the spectrum reduces to
\ber
I = \int_{-1}^{+1}\,du\,u\,\tanh(\Delta\,u/2T)
=
\Bigg\{
\begin{array}{ll}
	1                        &\,,\, T\rightarrow 0
	\cr
	\frac{\Delta(T)}{12 T_c} &\,,\, T\rightarrow T_c
\end{array} 
\,.
\eer
Note that near the transition the magnitude of the current decreases as $j_y\sim \Delta^2(T)/T_c\sim
(1-T/T_c)$, but also penetrates deeper into the bulk as
$\xi_{\text{$\Delta$}}=\hbar v_f/2\Delta(T)\sim (1-T/T_c)^{-\nicefrac{1}{2}}$.

\vspace*{-7mm}
\subsection{Edge Currents and Angular Momentum}\label{sec:edge_current}
\vspace*{-5mm}

Mass currents confined near the boundary (``edge currents'') generate macroscopic
angular momentum.
For a Galilean invariant system such as liquid \He\ the mass current density is obtained from the
spectral current density in Eq. \ref{JDOS} by the replacement $\vv_{\vp}\rightarrow m^* \vv_{\vp} =
\vp$, where $m^*$ is the quasiparticle effective mass, $\vp = p_f\hat\vp$ is the Fermi momentum.
In addition, $v_f$, $p_f$ and the normal-state density of states, $N_f$, determine the particle
number density, which for a 2D Fermi surface gives $n\equiv N/V = N_f\,p_f\,v_f$.

For a chiral p-wave superfluid confined within a thin cylindrical vessel of radius $R$ and height
$h$ in the 2D limit, $h\ll \xi_{\text{$\Delta$}}\ll R$, the angular momentum relative to the
$z$-axis is determined by the radial moment of the azimuthal component of the mass current density,
$g_{\varphi}(\vr)$,
\be\label{angular-momentum_cylinder}
L_z 
=
\int_{\text{V}} d^3r\,
\left[
r\,g_{\varphi}(\vr)
\right]
\,.
\ee
For $R\gg\xi_{\text{$\Delta$}}$ we can neglect the curvature of the surface, in which
case the azimuthal mass current is given by the tangential component of the boundary current
calculated from Eq. \ref{particle_current_density}.
Thus, the bound-state contribution to $L_z$ at $T=0$ obtained from Eqs.
\ref{angular-momentum_cylinder} and \ref{bound-state_current-density}, with $v_f\rightarrow p_f$
becomes,
\ber\label{Lz_bound-states}
L_z^{\text{bs}} 
&=&
N_f\,p_f\,\Delta\,\times\,2\pi\,h\int_{0}^{R} r^2\,dr\,e^{-(R-r)/\xi_{\text{$\Delta$}}}
=
N\,\hbar
\,,
\eer
which is a factor of \emph{two} larger than that predicted by Ishikawa \cite{ish77} and
McClure and Takagi\cite{mcc79} based on the real-space N-particle wave function of Eq.
\ref{BCS_wave-function}.
Finite size corrections from Eq. \ref{Lz_bound-states} are negligible - of order
$\xi_{\text{$\Delta$}}/R\lll 1$. As Stone and Roy pointed out the discrepancy is resolved by
including the contribution to $L_z$ from the states comprising the continuum spectrum.\cite{sto04}
Below I analyze the continuum contributions to the edge current and ground state angular
momentum.
In particular, I show that there are two contributions to the continuum spectral current density:
(i) an isolated scattering resonance that exactly cancels the bound-state contribution to the edge
current for each value of $\vp$ and (ii) a non-resonant response of the bound continuum that
accounts exactly for the MT result of $L_z = (N/2)\hbar$.

The energy range $\varepsilon<-\Delta$ constitutes the bound continuum, while the range
$\varepsilon>+\Delta$ represents excitations above the gap. At finite temperatures sub-gap surface
excited states $0 < \varepsilon < \Delta$ also play an important role. The spectral weight
associated with the continuum spectrum is modified near the boundary. For $|\varepsilon| > \Delta$,
$\lambda(\varepsilon) = i\,\sgn(\varepsilon)\sqrt{\varepsilon^2 - \Delta^2}$ and the spectral
function becomes,
\ber\label{DOS_continuum}
\cN_{\text{c}}(x,\vp;\varepsilon)
&=&
\frac{|\varepsilon|}{\sqrt{\varepsilon^2 - \Delta^2}}
\\
&-& 
\frac{|\varepsilon|}{\sqrt{\varepsilon^2 - \Delta^2}}
\frac{\Delta^2_{1}(\vp)}{\varepsilon^2-\Delta^2_2(\vp)}
\cos(2\sqrt{\varepsilon^2 - \Delta^2}\,x/v_x)
\nonumber
\\
&-&
\sgn(\varepsilon)\,
\left(
\frac{\Delta_{1}(\vp)\Delta_{2}(\vp)}{\varepsilon^2-\Delta^2_2(\vp)}
\right)
\sin(2\sqrt{\varepsilon^2 - \Delta^2}\,x/v_x)
\,.
\nonumber
\eer
The first term is the bulk continuum spectrum, while the corrections to the continuum spectrum 
are given by second and third lines in Eq. \ref{DOS_continuum}.
The third term is \emph{odd} under either $\varepsilon\rightarrow -\varepsilon$ or
$\vp\rightarrow -\vp$, and thus gives a non-vanishing contribution to the spectral current density,
\ber
\cJ_{\text{c}}(x,\vp;\varepsilon) 
&=&
-2N_f \vv_{\vp}\,\sgn(\varepsilon)
\left(
\frac{\Delta_{1}(\vp)\Delta_{2}(\vp)}{\varepsilon^2-\Delta^2_2(\vp)}
\right)
\nonumber\\
&\times&
\sin(2\sqrt{\varepsilon^2 - \Delta^2}\,x/v_x)
\eer
Note that for fixed energy $\varepsilon$ and momentum $\vp$ the effect of surface scattering on the
continuum spectrum is large and \emph{propagates} into the bulk. Thus, it is not a priori
clear that the current is confined to the surface. However, the wavelength of the disturbance is
given by the Tomasch wavelength for a specific trajectory,
\be
\lambda_{\text{T}}(\vp,\varepsilon) = \frac{\pi\hbar v_f\hat{p}_x}{\sqrt{\varepsilon^2 - \Delta^2}}
\,.
\ee
The net current parallel to the boundary is given by the sum over all incident trajectories,
\be\label{jy-continuuum}
j_y^{\text{c}}(x)
=
2N_f v_f
\,
\int_{-\pi/2}^{+\pi/2}\,\frac{d\alpha}{\pi}\,\hp_y\,\Delta_{1}(\vp)\Delta_{2}(\vp)
\times
J(\vp)
\,,
\ee
where $J(\vp)$ is given by 
\be\label{jy-continuuum-kernel}
J(\vp)
=
\int_{\Delta}^{\infty} \,d\varepsilon
\frac{\tanh(\varepsilon/2T)}{\varepsilon^2-\Delta^2_2(\vp)}
\times
\sin(2\sqrt{\varepsilon^2 - \Delta^2}\,x/v_x)
\,.
\ee
The integration over the spectral current density leads to phase cancellation away from the
boundary, and a \emph{net} current that is confined to the edge.
Although the integration is over the continuum spectrum, the chiral edge state nevertheless
modifies the current carried by the continuum states. Trajectories near grazing
incidence give a large enhancement to the kernel $J(\vp)$ coming from the off-resonant bound state.
The kernel is weighted by the product $\hp_{y}\,\Delta_1(\vp)\Delta_2(\vp)$, which is peaked
near $\alpha\approx\pm 55^{o}$.

At zero temperature the kernel is evaluated by transforming to an integration over 
the radial momentum $p$, or equivalently $\xi=v_f(p-p_f)$ with $\xi^2 = \varepsilon^2 - \Delta^2$,
\be
J(\vp)
=
\onehalf
\Im\,
\int_{\cC_{\text{R}}} 
\,d\xi
\frac{\xi}{(\xi^2+\Delta_1^2)\sqrt{\xi^2+\Delta^2}}
\times
e^{2i\xi\,x/v_x}
\,.
\ee
The singularities shown in the upper half of the complex $\xi$-plane determine the continuum current
response. In particular, the integral along the real axis is transformed to an integral around the
pole at $i|\Delta_{1}|$ and the branch cut from $i\Delta$ to $i\,\infty$: $J_{\cC_{\text{R}}} =
J_{\cC_{\text{1}}} + J_{\cC_{\text{2}}}$.
%
\begin{figure}[t]
\includegraphics[width=0.7\linewidth,keepaspectratio]{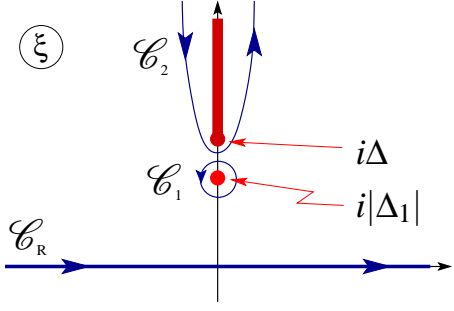}
\caption{\small
Integration in the complex $\xi$ plane. Integration along the real axis ($\cC_{\text{R}}$) is
transformed into the sum of an integral around the isolated pole at $i|\Delta_{1}|$ and the branch
cut along the imaginary $\xi$-axis.
\label{fig-complex_plane}
}
\end{figure}
%
The pole at $\xi=i|\Delta_{1}|$ is an isolated resonance that gives a contribution to the continuum
current that is confined to the boundary on the length scale $\xi_{\text{$\Delta$}}$,
\be
J_{\cC_{\text{1}}} = \frac{\pi}{2|\Delta_2(\vp)|}\,e^{-x/\xi_{\text{$\Delta$}}}
\,.
\ee
The current generated by this resonance exactly cancels the bound-state edge current
and bound-state contribution to the angular momentum,
\be
L_z^{\cC_{\text{1}}}
=
\int_{\text{V}} d^2r\,
\left[
r\,g_{\varphi}^{\cC_{\text{1}}}(\vr)
\right]
=
- N\hbar
\,.
\ee
Thus, the ground-state current and angular momentum come entirely from the 
non-resonant contribution to the continuum spectrum defined by the branch cut $\cC_{\text{2}}$ 
which evaluates to
\be
J_{\cC_{\text{2}}}
= 
-\int_{0}^{\infty}\frac{d\eps}{\eps^2+|\Delta_{2}(\vp)|^2}\,
e^{-2\sqrt{\eps^2+\Delta^2}\,x/v_x}
\,.
\ee
The current density is then given by
\ber
j_y^{\cC_{\text{2}}}(x)
&=&
2N_f v_f
\,
\int_{-\pi/2}^{+\pi/2}\,\frac{d\alpha}{\pi}\,\hp_y\,|\Delta_{1}(\vp)|\Delta_{2}(\vp)
\nonumber\\
&\times&
\int_{0}^{\infty}\frac{d\eps}{\eps^2+|\Delta_{2}(\vp)|^2}\,
e^{-2\sqrt{\eps^2+\Delta^2}\,x/v_x}
\,,
\label{chiral_edge_current}
\eer
which is confined to the edge, but in contrast to the bound-state and resonance terms, 
there is not a
single confinement length, but rather a weighted average of exponential confinement on length scales
$\pi\hbar v_f\,\cos\alpha/\Delta$. 
For this reason an analytic expression for the net current density analogous to Eq.
\ref{bound-state_current-density} does not appear possible. However, the total edge current and
ground state angular momentum can be computed by first carrying out the integration over the region
of the edge current. In the limit $R\gg\xi_{\text{$\Delta$}}$ the resulting ground-state angular
momentum reduces to the following integration over the continuum spectrum,
\ber\label{Lz_C2}
L_z^{\cC_{\text{2}}}
=
N\hbar &\times&
\frac{2}{\pi}
\int_{-\pi/2}^{+\pi/2}\,d\alpha\,p_x\,p_y\,\Delta_{1}(\vp)\Delta_{2}(\vp)
\\
&\times&
\int_{0}^{\infty}\frac{d\eps}{(\eps^2+|\Delta_2(\vp)|^2)
\sqrt{\eps^2+\Delta^2}}
\,,
\nonumber
\eer
which evaluates to (see Appendix \ref{appendix-angular_momentum_integration})
\be\label{Lz_MT-result}
L_z^{\cC_{\text{2}}} = \frac{N}{2}\hbar
\,.
\ee
Thus, Ishikawa\cite{ish77}, McClure and Takagi\cite{mcc79} and Stone and Roy's\cite{sto04} results
are recovered from the continuum response to the formation of the chiral edge state.

\vspace{-7mm}
\subsection{Temperature Dependence of $L_z$}\label{sec:Lz_vs_T}
\vspace{-5mm}

For $T\ne 0$ thermal excitations out of the ground state lead to a reduction of the order parameter
$\Delta(T)$, the edge current and angular momentum. The latter can be expressed as
\ber
L_z(T) =
\frac{N}{2}\hbar\,\times\cY_{\text{$L_z$}}(T) 
\,,
\eer
where $\cY_{\text{$L_z$}}(T)\rightarrow 1$ for $T\rightarrow 0$, vanishes for
$T\rightarrow T_c$, and can be calculated from the edge current at finite temperature.

Calculations of the temperature dependence of the angular momentum for \Hea\ were carried out by T.
Kita on the basis of numerical solutions to the Bogoliubov equations for mesoscopic cylindrical (3D)
geometries with dimensions $R\sim 4h\sim 2\xi$. Kita showed that $\cY_{\text{$L_z$}}(T)$ decreases
rapidly for $T\gtrsim 0$, indicating that there are low-lying excitations that are thermally
populated even at low temperatures which reduce the ground-state angular momentum.
Based on his numerical results (Fig. 2a of Ref. \onlinecite{kit98}), Kita conjectured that the
temperature dependence of $\cY_{\text{$L_z$}}(T)$ resulted from the excitations responsible for the
suppression of the superfluid density $\rho_{s,||}(T)$ of bulk \Hea\ corresponding to superflow
along the \emph{nodal direction} for the 3D chiral p-wave superfluid.
For 3D bulk superfluid \Hea\ the stiffness for $\vp_s ||\, \hvl$ is strongly suppressed at finite
temperature compared to the stiffness for superflow perpendicular to the nodal direction, i.e.
$\rho_{s,\perp} \gg \rho_{s,||}$.
However, as I discuss below, the softness of the angular momentum response function
$\cY_{\text{$L_z$}}(T)$ that Kita found numerically, including its near equality with
$\rho_{s,||}(T)$ for the 3D A-phase, is also present in the 2D limit in which the chiral p-wave
superfluid is fully gapped.

For $T\ne 0$ the edge current is determined by the continuum contribution to the
spectrum defined in Eq. \ref{jy-continuuum} with,
\be
J(\vp)
=
\onehalf
\Im\,
\int_{\cC_{\text{R}}} 
\,d\xi
\frac{\xi\,\tanh(\sqrt{\xi^2+\Delta^2}/2T)}{(\xi^2+\Delta_1^2)\sqrt{\xi^2+\Delta^2}}
\times
e^{2i\xi\,x/v_x}
\,.
\ee

\begin{figure}[t]
\includegraphics[width=0.8\linewidth,keepaspectratio]{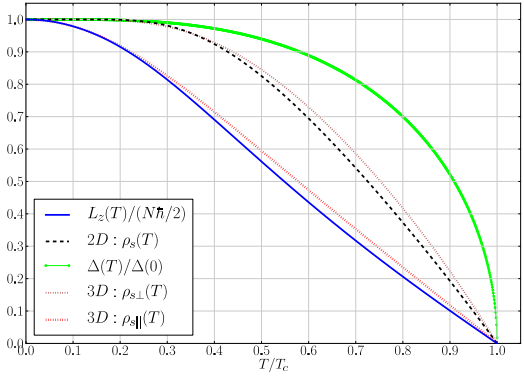}
\caption{\small 
Temperature dependence of the angular momentum, $L_z(T)$, for the 2D chiral p-wave superfluid
(\blue{\bf{--}}). Also shown is the superfluid stiffness (\black{\bf{--}}) and the bulk gap
(\green{\bf{--}}) for the fully gapped 2D chiral p-wave state. Shown for comparison are the two
components of the superfluid stiffness for the 3D chiral, p-wave superfluid phase (\Hea) -
$\rho_{s,||}$ for $\vp_s || \hvl$ (\red{\bf{--}}) and $\rho_{s,\perp}$ for $\vp_s\perp\hvl$
(\darkbrown{\bf{--}}).
\label{fig-Lz_vs_T}
}
\end{figure}

As is the case for $T=0$, the resonant contribution to the continuum current density coming
from the isolated pole at $\xi=+i|\Delta_{1}|$ exactly cancels the bound state contribution.
However, the total edge current and angular momentum, which at $T=0$ is calculated from the branch
cut in Fig. \ref{fig-complex_plane}, now results from the sum of contributions from a discrete set
of poles at the complex momenta defined by
\be
\xi_n = i\sqrt{\varepsilon_n^2 + \Delta^2}
\,,
\ee
where $\varepsilon_n=(2n+1)\pi T$, $n=0,\pm 1,\pm 2,\ldots$, are the fermion Matsubara
frequencies.\footnote{The transformation is based on the Matsubara representation for
$\tanh(\sqrt{\xi^2+\Delta^2}/2T)/2\sqrt{\xi^2+\Delta^2}
=T\sum_{\varepsilon_n}(\xi^2+\varepsilon_n^2+\Delta^2)^{-\nicefrac{1}{2}}$.}
The resulting edge current density is given by
\ber\label{continuum-current-finite_T}
j_y^{\cC_{\text{2}}}(x)
&=&
2N_f v_f
\,
\int_{-\pi/2}^{+\pi/2}\,\frac{d\alpha}{\pi}\,p_y\,|\Delta_{1}(\vp)|\Delta_{2}(\vp)
\nonumber\\
&\times&
\pi T\sum_{\varepsilon_n}
\frac{1}{\epsilon_n^2+|\Delta_{2}(\vp)|^2}
\,
e^{-2\sqrt{\epsilon_n^2+\Delta^2}\,x/v_x}
\,.
\eer
Multiple confinement scales are manifest in Eq. \ref{continuum-current-finite_T}. The total
surface current obtained by integrating over the boundary region determines the equilibrium
angular momentum generated by these edge currents,
\ber\label{Lz_vs_T} 
\cY_{\text{$L_z$}}(T)
=\frac{8}{\pi}\int_{0}^{1}dx\,(1-x^2)^{\tinyonehalf} 
\pi T\sum_{\varepsilon_n}
\frac{\Delta^2\,x^2}{\varepsilon_n^2+\Delta^2 x^2}
\frac{1}{\sqrt{\eps_n^2+\Delta^2}}
\,,
\eer
where $x=\hat{p}_y=\sin\alpha$. Figure \ref{fig-Lz_vs_T} shows the temperature dependence of the
equilibrium angular momentum, $\cY_{\text{$L_z$}}(T)$, calculated from Eq. \ref{Lz_vs_T}. Also shown
for comparison is the bulk excitation gap and superfluid stiffness for both 2D and 3D chiral p-wave
states. Note that the temperature dependence of the angular momentum is much softer than the bulk
superfluid stiffness for the gapless 2D phase.

Just as Kita found for his 3D mesoscopic geometry, the temperature dependence of $L_z(T)$ for the
fully gapped 2D phase is nearly identical to the superfluid stiffness for superflow parallel to the
nodal direction for the 3D phase. However, the reasons for the rapid suppression of $L_z(T)$ and
$\rho_{s,||}(T)$ are of different physical origin. For the bulk 3D phase $\rho_{s,||}(T)$ is
strongly reduced compared to $\rho_{s,\perp}(T)$ due to the backflow current carried by the nodal
excitations when $\vp_s ||\, \hvl$.\cite{xu94a} By contrast, for the fully gapped 2D chiral phase
there are low-energy backflow surface currents for $0 < \varepsilon < \Delta$ that reduce the edge
current when thermally populated (\emph{cf} Fig. \ref{fig-Jydos_bound-state}). The presence of
low-energy surface excitations is also evident in the spectral sums that define the edge current and
angular momentum in Eqs. \ref{continuum-current-finite_T}-\ref{Lz_vs_T}.

\vspace*{-7mm}
\subsection{Robustness of the Edge Currents}\label{sec:boundary_conditions}
\vspace*{-5mm}

The result of Ishikawa,\cite{ish77} and McClure and Takagi,\cite{mcc79}, for the ground-state
angular momentum is based a geometry with cylindrical symmetry, a chiral p-wave order parameter and
many-body wave function that is an eigenfunction of the angular momentum operator and two-particle
wave functions that vanish at the boundary. The analysis presented above relies on the
formation of edge states by boundary scattering in the presence of a chiral order parameter.
The resulting chiral edge states, and their dispersion relation shown in Fig.
\ref{fig-chiral_edge-state_dispersion}, play a key role in generating the edge current carried
by the continuum states and the resulting ground-state angular momentum of $(N/2)\hbar$.
One can ask ``how robust are these results to boundary conditions, geometry and topology?''

\begin{figure}[h]
\includegraphics[width=0.8\linewidth,keepaspectratio]{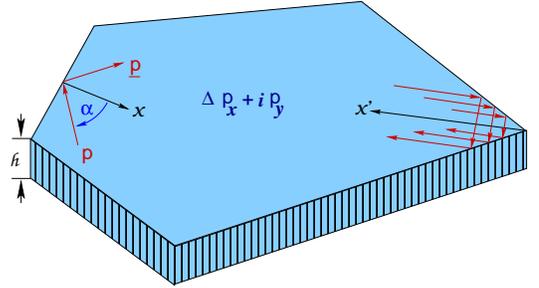}
\caption{\small
A thin film ($h\ll\xi$) of $p_x+ip_y$ superfluid (``2D \Hea'') confined in a non-cylindrical
geometry with area $\cA\ggg\xi^2$ bounded in the $x-y$ plane by specular surfaces. 
Double reflections are important in determining the surface
spectrum and edge current near a corner.
\label{fig-film_geometry}
}
\end{figure}

For example, consider the spectrum, edge currents and ground-state angular momentum for a geometry
such as that shown in Fig. \ref{fig-film_geometry}. There are two classes of trajectories
that determine the local spectral current density. Far from a corner ($\approx 5\xi$) trajectories
with a \emph{single reflection} determine the local surface spectrum, and for specular
reflections we obtain the chiral edge states and the local edge currents of Eqs.
\ref{chiral_bound-state} and \ref{chiral_edge_current}. However, near a corner the sharp change in
curvature leads to \emph{double reflections} as shown in Fig. \ref{fig-film_geometry}. These
double reflections dramatically alter the local excitation spectrum. 
They are also essential for enforcing current conservation near the corner, and they provide the
mechanism for the edge currents to ``turn the corner'' and maintain continuity of the current
circulating near the boundary. Furthermore,
since the double reflections are relevant only for incident trajectories within a few coherence
lengths of a corner the ground-state angular
momentum measured from the center of mass of the film is given by $(N/2)\hbar$ for a
finite number of corners, with corrections that are of order $\xi/\bar{R}\lll 1$, where
$\bar{R}$ is the minimum linear dimension of the film.

\begin{figure}[h]
\includegraphics[width=0.5\linewidth,keepaspectratio]{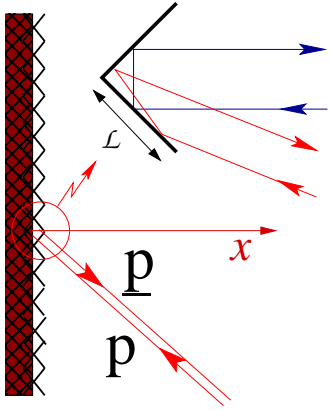}
\caption{\small
Mesoscopic facets of dimension $\hbar/p_f\ll\cL\ll\xi$ are retro-reflectors of quasiparticles.
\label{fig-retro-reflector}
}
\end{figure}

This example also indicates how non-specular scattering can dramatically alter the surface spectrum,
reduce or even eliminate the edge currents. To illustrate the effect non-specular scattering
consider a surface that is facetted with mesoscopic mirror surfaces that are large compared to the
Fermi wavelength, but small compared to the coherence length, $a \ll \cL \ll \xi$, and oriented at
right angles to one another as shown in Fig. \ref{fig-retro-reflector}. Such a surface is
a retro-reflector analogous to optical retro-reflectors constructed from dense packing of corner
reflectors.\cite{bar79}
Note that a retro-reflecting surface does not break time-inversion symmetry or reflection symmetry
in a plane containing the normal to the surface, but translational invariance is broken on
short-wavelength scales, $\cL\ll\xi$.
As a result, retro-reflection can dramatically modify the spectrum of edge
states.\footnote{Retro-reflective boundary scattering is not to be confused with Andreev reflection
- which is retro-reflection of the \emph{group velocity} associated with \emph{branch conversion}
between particle-like to hole-like excitations. Retro-reflective boundary scattering reverses the
momentum of the excitations. Andreev processes are also important in the superfluid phase, and
accounted for in the boundary condition for the Nambu propagator as described in Appendix
\ref{appendix-quasiclassical}.}
In the limit of \emph{perfect retro-reflection} - i.e. retro-reflection of all incident trajectories
- the spectrum of edge states is obtained by an analogous calculation to that of perfect specular
reflection since every incident trajectory is paired with a single reflected trajectory.
In particular, the quasiparticle propagator, and the corresponding bound-state spectral function,
are given by (see Appendix \ref{appendix-quasiclassical})
\ber\label{gR_retro-reflection}
\gR_{3}(x,\vp;\varepsilon) 
  &=&
-\frac{\pi\tepsR}
{\lambda(\varepsilon)}
+
\frac{\pi\Delta^2}{\lambda(\varepsilon)\tepsR}\,
      \,e^{-2\lambda(\varepsilon)x/v_x}
\,,
\\
\cN_{\text{bs}}(x,\vp;\varepsilon) 
  &=&
\pi|\Delta|\,e^{-2\Delta\,x/v_x}\,\delta(\varepsilon) 
\,.
\label{DOS_bound-state_retro-reflection}
\eer

In place of the chiral branch of edge states for perfect specular reflection (Fig.
\ref{fig-chiral_edge-state_dispersion}), perfect retro-reflection leads to an edge state at the
Fermi level, $\varepsilon_{\text{bs}}(\vp) = 0$, i.e. a \emph{zero-mode} for every incident
trajectory, $\vp$. These modes do not carry current, nor do they generate continuum currents. 
Indeed the spectral current density (Eq. \ref{JDOS}) vanishes identically, and thus
the ground-state angular momentum resulting from the edge states vanishes as well. 

The spectrum of zero-modes is also inferred from the observation that
$\Delta(\ul{\vp},x)=-\Delta(\vp,x)$ for any pair ($\ul{\vp},\vp)$ of retro-reflected trajectories.
Thus, Andreev's equation for a pair of retro-reflected trajectories is equivalent to Dirac fermions
in 1D coupled to a scalar field $\varphi(z)=\Delta\,\sgn(z)$ ($z$ being the coordinate measured
along the classical trajectory), which has the well-known Jackiw-Rebbi zero-mode
bound to the domain wall at $z=0$.\cite{jac76}

However, the zero modes generated by retro-reflection and a chiral p-wave order parameter are
fragile and unprotected from small perturbations. For an \emph{imperfect} retro-reflecting surface
some incident trajectories will be reflected \emph{forward} and generate edge
currents and a ground-state angular momentum with a magnitude in proportion to the probability for
forward reflection. Thus, depending on the distribution of trajectories with forward vs.
retro-reflection the resulting ground-state angular momentum will generally be less than
$(N/2)\hbar$, and may take on any value in the range, $L_{z}^{\text{min}} \lesssim L_z \le
(N/2)\hbar$, with the lower limit set by the \emph{intrinsic} angular momentum,\cite{vol75,cro77,bal85}
\be
L_{z}^{\text{min}}=(N/2)\hbar\times\nicefrac{1}{4}\,\left(\Delta/E_f\right)^2\ln(E_f/\Delta)
\,.
\ee

For \Hea\ confined by the walls of an experimental cell a realistic estimate for $L_z$ is likely
below $(N/2)\hbar$, but much larger than the intrinsic limit, and determined by the mean fraction
$f$ of \emph{forward} reflections by the boundary, i.e. $\sgn(p_y) = \sgn(\ul{p}_y$),
\be
L_z = f\,\times\,(N/2)\hbar
\,,
\ee
with $f^{\text{min}} \le f \le 1$.
The sensitivity of the ground-state angular momentum to retro-reflection is at first sight in
conflict with the result of McClure and Takagi (MT). However, the MT boundary condition does not
account for retro-reflection on mesoscopic scales because it assumes perfect cylindrical symmetry on
the atomic scale.
This result highlights the fact that spectrum of edge states, currents and the ground-state angular
momentum is sensitive to surface scattering on all scales from several coherence lengths down to the
atomic scale.

\begin{figure}[t]
\includegraphics[width=0.9\linewidth,keepaspectratio]{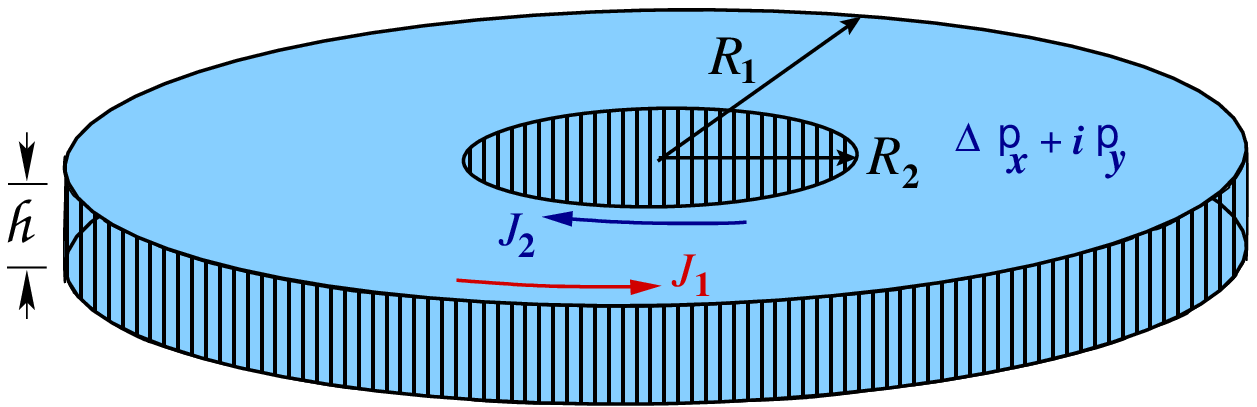}
\caption{\small
A thin film of chiral p-wave superfluid (``2D \Hea'') with $\mathit{h}\ll\xi$, inner and outer radii
$R_{2}$, $R_{1}$ and area $\cA =\pi(R_1^2-R_2^2)$ bounded by specular surfaces which reflect
quasiparticles $\vp\rightarrow\underline{\vp}=\vp-2\hvx(\hvx\cdot\vp)$. For $R_1,R_2,\, R_1-R_2 \gg
\xi$ only single reflections are relevant to determining the surface spectrum and edgecurrents on
the inner and outer boundaries.
\label{fig-toroidal_geometry}
}
\end{figure}

\vspace*{-7mm}
\subsection{Toroidal Geometry}\label{sec:toroidal_geometry}
\vspace*{-5mm}

The combination of geometry and surface boundary conditions can lead to dramatically different
results for the ground-state angular momentum of a chiral p-wave superfluid. Consider the toroidal
geometry shown in Fig. \ref{fig-toroidal_geometry} in which the superfluid is confined between inner
and outer baoundaries with radii $R_2$ and $R_1$, respectively. I assume both radii are large
compared to the confinement scale of the chiral edge currents, and that the edge states on the inner
and outer boundaries are well separated; i.e. $R_{1}, R_{2}, R_{1}-R_{2} \gg\xi_{\text{$\Delta$}}$.
The ground state angular momentum is given the radial moment of the mass current density in Eq.
\ref{angular-momentum_cylinder}, which in these limits is determined by the mass \emph{sheet
current} on the inner and outer boundaries, $K_2$ and $K_1$, respectively,
\ber
L_z = 2\pi\,h\,\left(K_1\,R_1^2 + K_2\, R_2^2\right)
\,.
\eer

At $T=0$ the magnitude of the mass sheet current (with units of ``action/volume'') for a specular
boundary is obtained from Eq. \ref{chiral_edge_current} with $v_f\rightarrow p_f$, and evaluates to
\be
K = \int_{0}^{\infty}dx\,g_{\varphi}(x) = \nicefrac{1}{4}\,N_f\,v_f\,p_f\,\hbar 
  = \frac{1}{4}\,n\,\hbar
\,.
\ee
For perfect specular reflection on both boundaries we obtain edge currents of equal magnitude
flowing in opposite directions, $K_1 = -K_2 = K$, as indicated in Fig.
\ref{fig-toroidal_geometry}, and thus once again the MT result for the ground-state angular
momentum,
\be\label{Lz_toroidal_specular}
L_z = 2\pi\,h\,\left(R_1^2 - R_2^2\right)\,\frac{1}{4}\,n\,\hbar = \frac{N}{2}\hbar
\,.
\ee
Note that the counter-propagating edge currents conspire to give a ground-state angular momentum, in
units of $\hbar/2$, that is extensive and proportional to the volume, or total number of particles.
If the boundary is not perfectly specular then the corresponding sheet current is reduced by the
suppression of the edge currents by retro-reflection: $K_f = f\times\nicefrac{1}{4}\,n\,\hbar$, with
suppression factor $ 0 < f < 1$.

For the toroidal geometry the inner and outer boundaries may have different degrees of specularity,
i.e. $K_1 = f_1\, K$ and $K_2 = -f_2\, K$ with $f_1 \ne f_2$. The generalization of Eq.
\ref{Lz_toroidal_specular} is
\be\label{Lz_toroidal_non-specular}
L_z = \frac{N}{2}\hbar\,\times \left(\frac{f_1 - r\,f_2}{1-r}\right)
\,,
\ee
where $1 < r \le 0$ is the ratio of the radii, $r=R_2/R_1$.
The asymmetry in the counter-propagating edge currents now leads to a ground-state 
angular momentum that no longer scales with the volume.
Two cases highlight the non-extensive property of $L_z$, and its sensitivity to the asymmetry in the
edge currents on different boundaries.

For perfect specular reflection on the outer boundary, $f_1=1$, and perfect retro-reflection on
the inner boundary, $f_2=0$, the resulting ground-state angular momentum 
\be
L_z = \frac{N}{2}\hbar\,\times \left(\frac{1}{1-r}\right)
\,,
\ee
can be much larger than the MT result $(N/2)\hbar$ for $1-r \ll 1$. 

Equally dramatic would be to engineer the outer boundary to be retro-reflecting, $f_1 = 0$, and the
inner boundary to be specular reflecting, $f_2 = 1$. In this limit only the counter-circulating
current on the inner boundary survives, which leads to a ground-state angular momentum that is
\underline{opposite} to the chirality of the Cooper pairs,
\be
L_z = \frac{N}{2}\hbar\,\times \left(\frac{-r}{1-r}\right)
\,.
\ee


This reversal of the ground-state angular momentum for a toroidal geometry would provide both
be a signature of the broken time-reversal symmetry of the ground state of superfluid \Hea, and
also establish its origin as the edge current from the inner boundary.

\vspace*{-7mm}
\subsection*{Acknowledgements}
\vspace*{-5mm}

This research is supported by the National Science Foundation (Grant DMR-0805277). I also
acknowledge the hospitality and support of the Aspen Center for Physics where part of this work was
carried out.

\vspace*{-7mm}
\section{Appendix: Boundary Solutions}\label{appendix-quasiclassical}
\vspace*{-5mm}

Using the representation for $\whgR$ in Eq. \ref{QCPropogator}, Eilenberger's equation can be
expressed as coupled equations for the quasiparticle and pair propagators in a three-dimensional
vector space,
\be\label{Vector_Eilenberger}
\nicefrac{1}{2}\,\vv_{\vp}\cdot\grad\ket{g} = \widehat{M}\ket{g}
\,,
\ee
with
\be
\ket{g}\equiv\begin{pmatrix}\fR_{1} \cr \fR_{2} \cr \gR_{3} \end{pmatrix}
\,,\quad
\widehat{M}
=
\begin{pmatrix}
	0 & \tilde\varepsilon^{R} & \Delta_{2} \cr
	- \tilde\varepsilon^{R} & 0 &-\Delta_{1} \cr
	\Delta_{2} & -\Delta_{1} & 0 
\end{pmatrix}
\ee

For a uniform order parameter defined by trajectory $\vp$ we express $\ket{g}$ in terms
of the eigenvectors of $\whM$, $\whM\ket{\mu}=\mu\ket{\mu}$. The eigenvector with $\mu=0$,
\be\label{egienvector0}
\ket{0;\vp} = \frac{1}{\lambda(\vp,\varepsilon)}
\begin{pmatrix}-\Delta_{1}(\vp) \cr -\Delta_{2}(\vp) \cr +\tepsR \end{pmatrix}
\,,
\ee
generates the bulk equilibrium propagator, 
\be
\whgR_{0} = -\frac{\pi}{\lambda}\left(\tepsR\whtauz - \whDelta(\vp)\right)
\,,
\ee
where $\lambda=\sqrt{|\Delta(\vp)|^2 - (\tepsR)^2}$ and
$|\Delta(\vp)|^2 = \Delta_{1}^2(\vp)+\Delta_{2}^2(\vp)=\Delta^2$.
This solution satisfies Eilenberger's normalization condition in Eq. \ref{Normalization_condition}.
There is also a pair of eigenvectors with eigenvalues $\mu=\pm\lambda$
\be\label{egienvector+-}
\ket{\pm;\vp} = \frac{1}{\sqrt{2}\lambda\lambda_{1}}
\begin{pmatrix}
	\pm\lambda\tepsR-\Delta_{1}\Delta_{2} \cr 
	\lambda_{1}^{2} \cr 
	\tepsR\Delta_{2}\mp\lambda\Delta_{1}
\end{pmatrix}
\,,
\ee
with $\lambda_{1}\equiv\sqrt{\Delta_{1}^2(\vp) - (\tepsR)^2}$.
These eigenvectors generate ``exploding solutions'' to Eq. \ref{Vector_Eilenberger} for energies
within the gap of the bulk quasiparticle spectrum, $|\varepsilon| < |\Delta(\vp)|$, and thus are
physical solutions only in the vicinity of a boundary, or near a localized
defect such as a vortex or domain wall.\cite{thu84}
For the same value of momentum, $\vp$, the eigenvectors are orthonormal, 
$\braket{\mu;\vp}{\nu;\vp}=\delta_{\mu\nu}$.\footnote{The eigenvectors
$\bra{\mu;\vp}$ are obtained from the adjoint of $\ket{\mu;\vp}$ and the replacement
$\tepsR\rightarrow-\tepsR$ since $\whM^{\dag}(\tepsR)=\whM(-\tepsR)$.}
The Nambu propagators corresponding to the eigenvectors $\ket{\pm,\vp}$ are
\ber
\whgR_{\pm}(\vp,\varepsilon)
=
\frac{1}{\sqrt{2}\lambda\lambda_{1}}
&\Big(&
(\tepsR\Delta_{2}\mp\lambda\Delta_{1})\,\whtauz 
\nonumber\\
&\mp&
i\sigma_x\,(\lambda\tepsR\mp\Delta_{1}\Delta_{2})\,\whtauy	
\nonumber\\
&+&
i\sigma_x\,\lambda_{1}^{2}\,\whtaux	
\Big)
\,.
\eer
These matrices are non-normalizable and anti-commute with the bulk propagator,
\be\label{Nambu_Algebra}
\left(\whgR_{\pm}\right)^2 = 0
\,\,,\qquad
\commutator{\whgR_{0}}{\whgR_{\pm}}_{+} = 0
\,.
\ee

For a boundary far from other boundaries or defects we must exclude solutions that explode
into the bulk of the superfluid.
In particular, for a pair of specular or retro-reflected trajectories the solutions for the incident
and reflected trajectories are
\ber
\ket{g_{\text{in}}(\vp,x)}
&=&
\ket{0;\vp} + C_{\text{in}}(\vp)\,e^{-2\lambda(\vp,\varepsilon)x/v_{x}}\ket{+;\vp}
\,,\\
\ket{g_{\text{out}}(\ul{\vp},x)}
&=&
\ket{0;\ul{\vp}} + 
C_{\text{out}}(\ul{\vp})\,e^{-2\lambda(\ul{\vp},\varepsilon)x/v_{x}}\ket{-;\ul{\vp}}
\,,
\eer
where $v_x = v_f\cos(\alpha)$ for $-\pi/2<\alpha<\pi/2$ and $x\ge 0$ is the coordinate normal to the
boundary as shown in Fig. \ref{fig-time-reversed_trajectories}. 
The corresponding Nambu propagator for the incident trajectory in the vicinity of the boundary is
constructed from these solutions with Eqs. \ref{Normalization_condition} and \ref{Nambu_Algebra} to
fix the normalization,
\be\label{Nambu_incident}
\whgR_{\text{in}} 
= -\pi\,
\Big(
\whgR_{0}(\vp,\varepsilon)
+
C_{\text{in}}(\vp,\varepsilon)\,\whgR_{+}(\vp,\varepsilon)\,e^{-2\lambda(\varepsilon)\,x/v_x}
\Big)
\,.
\ee

\vspace*{-7mm}
\subsection{Specular Reflection}
\vspace*{-5mm}

For an incident trajectory $\vp=(p_x,p_y)$, the specularly reflected trajectory is
${\ul{\vp}}=(-p_x,p_y)$. Thus, the eigenvectors for the specularly
reflected trajectory $\ul\vp$ are obtained from Eqs. \ref{egienvector0} and \ref{egienvector+-} by
the replacement, $\Delta_{1}\rightarrow-\Delta_{1}$.
The specular boundary condition requires continuity of the incoming and outgoing propagators at
$x=0$, which fixes the amplitudes, $C_{\text{in}}(\vp,\varepsilon)$ and
$C_{\text{out}}(\ul\vp,\varepsilon)$. For the incident trajectory,
\ber\label{C_amplitude_specular}
C^{\text{spec}}_{\text{in}}(\vp) = \frac{1-\braket{0;\ul\vp}{0;\vp}}{\braket{0;\ul\vp}{+;\vp}}
                                 = \frac{\sqrt{2}\Delta_1(\vp)\lambda_{1}(\vp,\varepsilon)}
                                   {\lambda(\varepsilon)\tepsR - \Delta_{1}(\vp)\Delta_{2}(\vp)}
\,.
\eer
Similarly for the specularly reflected trajectory:
$C^{\text{spec}}_{\text{out}}(\ul\vp,\varepsilon) = C^{\text{spec}}_{\text{in}}(\vp,\varepsilon)$.
The resulting propagator from Eqs. \ref{Nambu_incident} and \ref{C_amplitude_specular} gives the
results for the pair propagators, $\fR_{1,2}$, and quasiparticle propagator $\gR_{3}$ in Eqs.
\ref{fR1}-\ref{gR3}.

\vspace*{-7mm}
\subsection{Retro-Reflection }
\vspace*{-5mm}

For retro-reflection we have $\ul\vp=(-p_x,-p_y)$, and in this case the eigenvectors are obtained
from Eqs. \ref{egienvector0} and \ref{egienvector+-} by the replacements,
$\Delta_{1}\rightarrow-\Delta_{1}$ and $\Delta_{2}\rightarrow-\Delta_{2}$. This boundary condition
dramatically alters the propagator near the boundary with
\ber
C^{\text{retro}}_{\text{in}}(\vp) = \frac{\sqrt{2}\,\Delta^2\lambda_{1}(\vp,\varepsilon)}
  {\tepsR\,\left(\lambda(\varepsilon)\Delta_{1}(\vp) - \tepsR\Delta_{2}(\vp)\right)}
\,,
\eer
which gives the propagator for retro-reflection in Eq. \ref{gR_retro-reflection}, with a spectrum of 
zero-modes replacing the branch of chiral edge states for specular reflection.

\vspace*{-7mm}
\section{Appendix: Angular Momentum Integration}\label{appendix-angular_momentum_integration}
\vspace*{-5mm}

The second integral in Eq. \ref{Lz_C2}, derived from the branch cut in Fig. \ref{fig-complex_plane},
evaluates to
\be
\int_{0}^{\infty}\frac{d\eps}{(\eps^2+|\Delta_2(\vp)|^2)\sqrt{\eps^2+\Delta^2}}
=
\frac{1}{|\Delta_1|\,|\Delta_2|}\,\tan^{-1}\left(\frac{|\Delta_1|}{|\Delta_2|}\right)
\,.
\ee
Setting $|\Delta_{2}|=\Delta\,t$, $|\Delta_{1}|=\Delta\,\sqrt{1-t^2}$ reduces Eq. \ref{Lz_C2} to 
\ber
L_z^{\cC_{\text{2}}}
=
N\hbar &\times&
\frac{4}{\pi}\,\times\,\int_{0}^{1}\,dt\,\,t\,\tan^{-1}\left(\frac{\sqrt{1-t^2}}{t}\right)
\,.
\eer
Integration by parts reduces to a Beta function,\cite{abramowitz70}
\ber
\int_{0}^{1}\,dt\,\,t\,\tan^{-1}\left(\frac{\sqrt{1-t^2}}{t}\right)
&=&
\frac{1}{4}\,\cB(\nicefrac{3}{2},\nicefrac{1}{2})
=
\frac{\pi}{8}
\,,
\eer
which yields the MT result, $L_z=(N/2)\hbar$, given in Eq. \ref{Lz_MT-result}.

%

\end{document}